\documentclass{Interspeech}

\interspeechcameraready

\title{STOPA: A Dataset of \texorpdfstring{\underline{S}}{S}ystematic Varia\texorpdfstring{\underline{T}}{T}ion \texorpdfstring{\underline{O}}{O}f Dee\texorpdfstring{\underline{P}}{P}fake \texorpdfstring{\underline{A}}{A}udio for Open-Set Source Tracing and Attribution}

\author[affiliation={1}]{Anton}{Firc}
\author[affiliation={2}]{Manasi}{Chhibber}
\author[affiliation={2}]{Jagabandhu}{Mishra}
\author[affiliation={2}]{Vishwanath Pratap}{Singh}
\author[affiliation={2}]{Tomi H.}{Kinnunen}
\author[affiliation={1}]{Kamil}{Malinka}

\affiliation{}{Brno University of Technology}{Czech Republic}
\affiliation{}{University of Eastern Finland}{Finland}
\email{ifirc@fit.vut.cz, manasi.chhibber@uef.fi, jagabandhu.mishra@uef.fi, vsingh@uef.fi, tomi.kinnunen@uef.fi, malinka@fit.vut.cz}

\usepackage{comment}
\usepackage{pifont}
\usepackage{multirow}

\begin{document}

\maketitle

% the abstract here must exactly match the abstract entered into the paper submission system
\begin{abstract}
    A key research area in deepfake speech detection is source tracing — determining the origin of synthesised utterances. The approaches may involve identifying the acoustic model (AM), vocoder model (VM), or other generation-specific parameters. However, progress is limited by the lack of a dedicated, systematically curated dataset. To address this, we introduce STOPA, a systematically varied and metadata-rich dataset for deepfake speech source tracing, covering 8 AMs, 6 VMs, and diverse parameter settings across 700k samples from 13 distinct synthesisers. Unlike existing datasets, which often feature limited variation or sparse metadata, STOPA provides a systematically controlled framework covering a broader range of generative factors, such as the choice of the vocoder model, acoustic model, or pretrained weights, ensuring higher attribution reliability. This control improves attribution accuracy, aiding forensic analysis, deepfake detection, and generative model transparency.
\end{abstract}

\section{Introduction}

The proliferation of deepfakes has led to increasingly realistic synthetic speech, posing significant threats to humans and speaker recognition. Malicious actors exploit deepfake speech for impersonation or fraud, with AI-generated voices even used to initiate fraudulent financial transactions~\cite{deepfakeSurvey}.

Various detection approaches, from handcrafted feature-based to deep learning, have been developed to counter these threats~\cite{WANG2020101114, Wang2024}.
While substantial progress has been made, they  
%these methods effectively identify synthetic speech, they often 
provide limited insight into \emph{how} or \emph{who} generated given speech.
% the speech was generated or who created it. 
%This has driven interest in \textit{source tracing}. 
Unlike deepfake detection, which determines whether an utterance is real or fake, \textbf{source tracing} (or attack attribution) assumes this classification has already been made. The aim is to identify the entity or system responsible for generating the speech. 
% Source tracing 
These methods thus offer deeper insights into the generation process, thereby enhancing forensic investigations~\cite{chhibber2024explainableprobabilisticattributeembedding, yan-source-tracing}.

Despite recent progress, the lack of specialized datasets limits source tracing research. Existing datasets (reviewed in Section~\ref{sec:related-datasets}) focus on detection but lack \textbf{systematic variation} in acoustic models, vocoders, and other generative parameters. For example, the widely-adopted ASVspoof %challenge 
dataset series~\cite{WANG2020101114, Wang2024, ASVspoof21-dataset} is collected through semi-controlled crowdsourcing, where attack crafting (i.e. text-to-speech or voice conversion system development) follows a common generation protocol---but with %are created using diverse 
lot of flexibility %allowed on 
methods and their implementations, chosen by %the individual 
the data contributors. 
%Despite providing large-scale data resources for training and comparing detectors,  
%ensures state-of-the-art synthesis methods, 
%the variability in vocoders and acoustic models makes direct comparisons difficult. 
Although this design suits deepfake detection tasks, it prevents systematic analysis of attack characteristics essential for source tracing.

% Despite progress, a lack of specialized datasets slows the development of source tracing techniques. Existing datasets (reviewed briefly in Section~\ref{sec:related-datasets}) focus on detection and lack systematic variation in acoustic models, vocoders, and other generative parameters, limiting advances in source tracing research. For instance, the ASVspoof dataset sereies~\cite{WANG2020101114, ASVspoof21-dataset}, including the latest edition~\cite{Wang2024}, have been collected through a semi-controlled crowd-sourcing approach. While the spoofing attacks are based on common generation protocol, they have been crafted by individual data contributors leveraging techniques of interest to them. Consequently, despite the generated attacks representing state-of-the-art synthesis technology, the architectural choices on vocoders and acoustic models are widely varied, making the synthesis approaches not easily compared. While such a design is well-motivated to address the detection task, a systematic comparison of the traits that differentiate the attacks is impossible.

To address this research gap, we introduce \textbf{STOPA}\footnote{\texttt{https://doi.org/10.5281/zenodo.15462919}}
% \footnote{\textcolor{magenta}{Note to Interspeech 2025 reviewers: this footnote will be populated by a data repository link, conditional on paper acceptance. The supplementary ZIP file submitted with our paper provides a snapshot of the planned dataset format.}}
, a new dataset specifically designed for source tracing in deepfake audio detection. Our dataset features a systematic variation of key generative components, allowing researchers to analyze the impact of different model choices. 
% https://doi.org/10.5281/zenodo.15462920
% For example, we include four distinct attacks generated using the Tacotron2~\cite{shenTacotron} acoustic model with the following properties:
For example, we include four distinct attacks generated using Tacotron2~\cite{shenTacotron}:
\begin{itemize}
    \item Pairs of attacks share the same Tacotron2 implementation but differ in their vocoder choice.
    \item Tacotron2 models vary in architecture parameters (e.g., the number of attention heads) and pretrained weights.
\end{itemize}

While much of the recent focus has been on newer text-to-speech (TTS) models, %older 
legacy systems like Tacotron2 remain relevant due to their stable implementations, widespread availability, and ability to produce high-quality speech~\cite{gonzalezdocasal22_iberspeech}. If detection systems fail to handle these well-established models, they present an attractive option for attackers.

% To demonstrate the usage of our dataset, inspired by speaker detection, we tackle the source tracing problem as a detection task. A hypothesized spoofing attack, vocoder model or acoustic model is tested against an unlabeled test instance. We first train an embedding extractor that generates unique embeddings for known attacks, then compare test utterances against these embeddings for source identification. This leads to a significant number of test trials that can accordingly be divided into well-defined positive (same attack/vocoder/acoustic model) and negative (different attack/vocoder/acoustic models) tests following the trial ground truth. A fundamental principle adapted from speaker detection research is the need for a continuously expanding database rather than a statically predefined one. The proposed protocol is explicitly designed to support an evolving dataset, enabling the incorporation of new attack signatures over time without necessitating computationally expensive retraining of system components. This approach enhances the system's practicality and scalability, allowing for more efficient and adaptable source-tracing methods.

\textbf{A key difference of STOPA compared to existing datasets and most of the published methodology literature relies on the
% design of its evaluation protocols, 
evaluation protocols design,
intended to promote % for promoting 
development of source tracing methods for `open-world' settings.} To elaborate, the majority of the source-tracing literature %primarily 
approaches the task as a multiclass task %classification problem
~\cite{chhibber2024explainableprobabilisticattributeembedding, yan-source-tracing, Borrelli2021, bhagtani-source-tracing, bartusiak2022transformerbasedspeechsynthesizerattribution} assuming known attacks during training and handling unknown attacks via an additional fallback class with a dedicated out-of-distribution classifier.
% where the attacks (deepfake algorithms) are either assumed to be known at the training stage, or where the unknown attacks are treated as an additional fallback class, treated usually as an additional task with a dedicated 'out-of-distribution' type classifier. 
Whereas this treatment leads to 
% the dependency of typical performance metrics on
class priors dependencies in performance metrics, %the design of additional classifiers 
% adding 
extra classifiers further complicate system design and evaluation.

Inspired by the established principles of the speaker detection evaluation benchmarks coordinated by NIST~\cite{GREENBERG2020101032}, we frame source tracing as an \textbf{open-world detection task} that unifies source tracing and 'detecting the unknown' under a simple and coherent evaluation framework. A test instance is evaluated against a hypothesized spoofing attack, vocoder, or acoustic model. The task is to answer whether or not the test instance originates from the hypothesized source generator (null hypothesis) or not (alternative hypothesis). Similar to NIST SREs, %which are 
designed to support 'continual growing' of a speaker database, our %evaluation 
protocol forbids the use of %the 
\emph{other} attacks for training or testing purposes: a decision must solely be made by comparing the hypothesized model against a single test instance. This %change in evaluation 
%evaluation design 
%overall mindset %change is 
mindset intends to promote the design of %zero-shot 
deepfake profiling approaches where the database of deepfake 'signatures' (similar to databases of computer viruses or malware) is allowed to grow dynamically during the lifespan of the profiling system. A 'fixed' dataset of deepfake generators, addressed as a multiclass task, requires re-training the profiling system whenever a new deepfake generator is encountered.

\textbf{STOPA's main contributions include}: \textit{(1)} a scalable evaluation protocol enabling open-world source tracing by integrating new attack signatures without retraining, \textit{(2)} systematic variation in spoofed speech with diverse acoustic models, vocoders, and hyperparameters, and \textit{(3)} extensive metadata for fine-grained analysis and benchmarking of synthetic speech.

The benchmark results and code base are available at \texttt{https://github.com/Manasi2001/STOPA}.

% \begin{enumerate}
%     \item Systematic variation in generating spoofed speech, covering multiple acoustic models, vocoders, and hyperparameters.
%     \item Extensive metadata, providing detailed insights into the generative pipeline for further analysis and benchmarking.
% \end{enumerate}

\begin{table}[t]
    \centering
    \caption{Overview of related datasets. Column \textit{Var.} indicates whether the dataset systematically varies acoustic models (AMs) and vocoders (VMs) to facilitate source tracing.}
    \label{tab:related-datasets}
    \resizebox{\columnwidth}{!}{
    \begin{tabular}{lrrrrrrc}
        \toprule
        \textbf{Dataset} & \textbf{\#Utt.} & \textbf{\#Sys.} & \textbf{\#Spk.} & \textbf{\#AM} & \textbf{\#VM} & \textbf{Var.} & \textbf{Avail.} \\
        \midrule
        ASVspoof19 LA~\cite{WANG2020101114} & 121K & 19 & 107 & 19 & 11 & \ding{55} & Pub. \\
        ASVspoof21 DF~\cite{ASVspoof21-dataset} & 593K & 100+ & 93 & 19+ & 11+ & \ding{55} & Pub. \\
        ASVspoof5~\cite{Wang2024}  & 1.2M & 32 & 1,922 & 22 & N/A & \ding{55} & Pub. \\
        SemaFor~\cite{semafor-dataset}  & 17K & 11 & 25 & 11 & N/A & \ding{55} & Pub. \\
        MLAAD~\cite{muller2024mlaad}  & 154K & 82 & N/A & 13 & 6 & \ding{55} & Pub. \\
        TIMIT-TTS~\cite{timit-tts}  & 79K & 12 & 37 & 12 & 2 & \ding{55} & Pub. \\
        \cite{AlBadawy_2019_CVPR_Workshops}  & 1.8K & 5+ & 9 & N/A & N/A & \ding{55} & Int. \\
        \cite{yan-source-tracing}  & 63K & 8 & 692 & 1 & 8 & \ding{55} & Int. \\
        \cite{zhang2024distinguishingneuralspeechsynthesis}  & N/A & 17 & N/A & 3 & 14 & \ding{51} & Int. \\
        \midrule
        \textbf{STOPA (ours)} & \textbf{699K} & \textbf{13} & \textbf{107} & \textbf{8} & \textbf{6} & \textbf{\ding{51}} & \textbf{Pub.} \\
        \bottomrule
    \end{tabular}
    }
    \label{tab:datasets}
\end{table}

\section{Related work}

\subsection{Speech synthesis and voice conversion}

% The usual two-stage pipeline for speech synthesis includes an acoustic model (AM) for spectrogram generation and a vocoder model (VM) to convert the spectrogram into an audible waveform~\cite{DiffuseOrConfuse-dataset}. 
Speech synthesis typically follows a two-stage pipeline: an acoustic model (AM) generates a spectrogram, which a vocoder model (VM) converts into an audible waveform~\cite{DiffuseOrConfuse-dataset}.
AMs fall into two main categories: text-to-speech (TTS) and voice conversion (VC)~\cite{deepfakeSurvey}. TTS models transform the text into a speech representation. Relevant AMs include autoregressive models (e.g., Tacotron2~\cite{shenTacotron}), non-autoregressive models (e.g., FastPitch~\cite{fastpitch}), flow-based models (e.g., GlowTTS~\cite{casanova21b_interspeech,Glow-TTS}), and diffusion-based models (e.g., DiffGAN-TTS~\cite{liu2022diffganttshighfidelityefficienttexttospeech}).

% Relevant architectures include Tacotron2~\cite{shenTacotron}, a popular TTS architecture utilizing encoder-decoder architecture to directly predict mel-spectrograms from input text. DiffGAN-TTS~\cite{liu2022diffganttshighfidelityefficienttexttospeech} is a Denoising Diffusion Probabilistic model (DDPM)-based model that incorporates a shallow diffusion mechanism to reduce the amount of denoising steps required to generate mel-spectrograms. GlowTTS~\cite{casanova21b_interspeech,Glow-TTS} is a flow-based generative model that employs normalising flows to enable efficient and parallel mel-spectrogram synthesis. FastPitch~\cite{fastpitch} is a non-autoregressive model incorporating explicit pitch contours through a learned pitch predictor conditioned on phoneme duration.

In contrast, VC algorithms generate speech by combining linguistic content from one input with the vocal characteristics of another. These models may use representation learning techniques (e.g., SpeechSplit~\cite{pmlr-v119-qian20a}) or diffusion-based methods (e.g., Diff-VC~\cite{popov2022diffusionbasedvoiceconversionfast}).
% For example, SpeechSplit~\cite{pmlr-v119-qian20a} is a disentangled speech representation learning model that decomposes speech into pitch, rhythm, and content, making it effective for voice conversion. Diff-VC~\cite{popov2022diffusionbasedvoiceconversionfast} is a diffusion-based voice conversion model that uses a conditional DDPM with a Transformer-based denoiser to map source speech features to target mel-spectrograms.

Most commonly used vocoders include rule-based approaches (e.g., Griffin-Lim~\cite{griffinLim}), GAN-based methods (e.g., MelGAN~\cite{melgan}, HiFi-GAN~\cite{hifigan}), and diffusion-based models (e.g., WaveGrad~\cite{chen2020wavegradestimatinggradientswaveform}).

% Ultimately, notable VMs include WaveGrad~\cite{chen2020wavegradestimatinggradientswaveform}, a diffusion-based vocoder that generates waveforms by iteratively refining noise using learned gradients. GriffinLim~\cite{griffinLim}, a traditional phase-reconstruction algorithm that estimates phase information from a given mel-spectrogram through iterative refinement. HiFi-GAN~\cite{hifigan}, a high-fidelity generative adversarial network-based vocoder that efficiently synthesizes speech waveforms. MelGAN~\cite{melgan}, a GAN-based vocoder designed for real-time speech synthesis by leveraging a fully convolutional architecture with multi-scale discriminators. ParallelWaveGAN~\cite{pwgan}, a non-autoregressive vocoder using a combination of GAN and WaveNet-based approaches to generate waveforms in a parallel and efficient manner.

\subsection{Deepfake audio datasets}
\label{sec:related-datasets}

The most widely recognized and adapted datasets for anti-spoofing are the ASVspoof datasets~\cite{WANG2020101114, Wang2024, ASVspoof21-dataset}. Additionally, MLAAD~\cite{muller2024mlaad}, TIMIT-TTS~\cite{timit-tts}, and SemaFor~\cite{semafor-dataset} provide additional synthetic speech collections. A comparison with related datasets is shown in Table~\ref{tab:related-datasets}. While effective for anti-spoofing, these datasets were not designed for source tracing. However, some have been repurposed for this task.

The Interspeech 2025 special session on source tracing\footnote{\texttt{https://deepfake-total.com/sourcetracing}} uses MLAAD~\cite{muller2024mlaad} dataset as a benchmark, incorporating a source tracing protocol. MLAAD includes 38 languages and samples generated from 82 different synthesis systems spanning 38 architectures, ranging from traditional models like Tacotron2 and Griffin-Lim to state-of-the-art approaches such as XTTS and VITS. The dataset introduces variation through the language dimension, enabling analysis of synthesis methods across different training corpora and model configurations. The baseline system~\cite{xie24_interspeech} reports a 63\% EER, providing a reference for source tracing performance on large-scale multilingual datasets.

Additionally, several studies address source tracing using existing datasets. ASVspoof19 LA has been utilised in~\cite{chhibber2024explainableprobabilisticattributeembedding, Borrelli2021, bhagtani-source-tracing}, SemaFor in~\cite{bhagtani-source-tracing, bartusiak2022transformerbasedspeechsynthesizerattribution}, MLAAD in~\cite{klein24_interspeech}, and other researchers~\cite{yan-source-tracing, AlBadawy_2019_CVPR_Workshops, zhang2024distinguishingneuralspeechsynthesis} created and used internal datasets.
% STOPA introduces a systematic variation of acoustic and vocoder models while maintaining controlled overlap, enabling more rigorous experimentation. It provides detailed metadata on model settings and synthesis parameters, allowing for in-depth analysis of extractable information from synthetic speech. By structuring the data in this way, our dataset enables an alternative approach to source tracing that complements existing datasets, offering new possibilities for attribution research.
STOPA thus complements existing datasets by introducing systematic variation in acoustic models, vocoders, and hyperparameters, enabling fine-grained analysis. Unlike prior works, it employs an open-world evaluation protocol, %making source tracing more scalable and realistic.
as a step towards solutions expected in audio forensics and other applications.

\begin{figure}[t]
    \centering
    \includegraphics[width=0.75\linewidth]{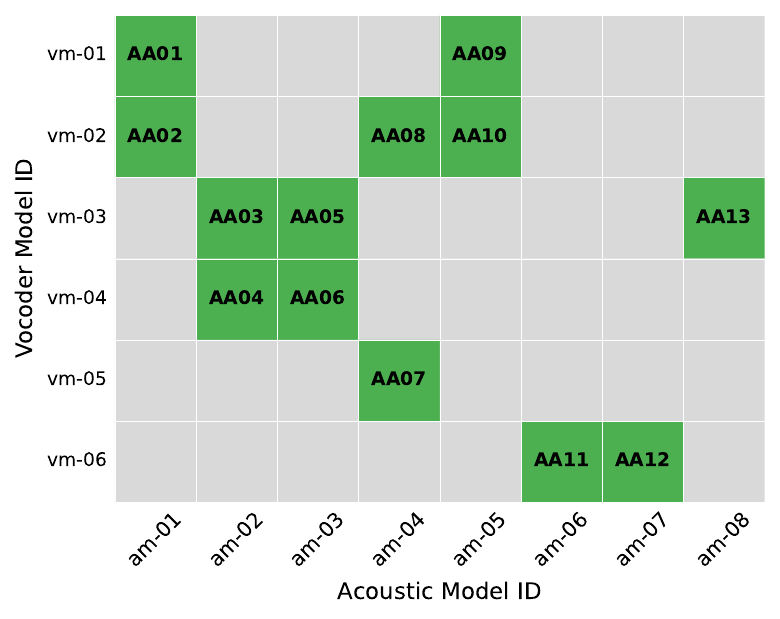}
    \caption{Combination of AMs and VMs for attacks}
    \label{fig:attacks}

    \vspace{1em}

    \captionof{table}{Used model architectures and implementations}
    \label{tab:used-models}
    \resizebox{0.85\columnwidth}{!}{
    \begin{tabular}{lllc}
    \toprule
    \textbf{ID} & \textbf{Type} & \textbf{Architecture} & \textbf{Implementation} \\
    \midrule
    \textit{am-01} & TTS & Tacotron2~\cite{shenTacotron} & \cite{Eren_Coqui_TTS_2021} \\
    \textit{am-02} & TTS & Tacotron2~\cite{shenTacotron} & \cite{lee2021comprehensive-tacotron2} \\
    \textit{am-03} & TTS & DiffGAN-TTS\cite{liu2022diffganttshighfidelityefficienttexttospeech} & \cite{Lee_DiffGAN-TTS_2022} \\
    \textit{am-04} & TTS & GlowTTS~\cite{casanova21b_interspeech}  & \cite{Eren_Coqui_TTS_2021} \\
    \textit{am-05} & TTS & FastPitch~\cite{fastpitch} & \cite{Eren_Coqui_TTS_2021} \\
    \textit{am-06} & TTS & GlowTTS~\cite{Glow-TTS} & \cite{glowTTS} \\
    \textit{am-07} & VC & SpeechSplit~\cite{pmlr-v119-qian20a} & \cite{speechSplit} \\
    \textit{am-08} & VC & Diff-VC~\cite{popov2022diffusionbasedvoiceconversionfast} & \cite{diff-vc} \\
    \midrule
    \textit{vm-01} & VM & WaveGrad~\cite{chen2020wavegradestimatinggradientswaveform} & \cite{Eren_Coqui_TTS_2021} \\
    \textit{vm-02} & VM & GriffinLim~\cite{griffinLim} & \cite{Eren_Coqui_TTS_2021} \\
    \textit{vm-03} & VM & HiFi-GAN~\cite{hifigan} & \cite{lee2021comprehensive-tacotron2} \\
    \textit{vm-04} & VM & MelGAN~\cite{melgan} & \cite{melgan-neurips} \\
    \textit{vm-05} & VM & HiFi-GAN~\cite{hifigan} & \cite{Eren_Coqui_TTS_2021} \\
    \textit{vm-06} & VM & ParallelWaveGAN~\cite{pwgan} & \cite{pwgan-sw} \\
    \bottomrule
    \end{tabular}}

\end{figure}

\section{STOPA dataset description}

\subsection{Creation procedure}

We introduce a new dataset for synthetic speech source tracing, systematically varying acoustic and vocoder models across multiple two-stage speech synthesis tools (Figure~\ref{fig:attacks}, Table~\ref{tab:used-models}). Unlike end-to-end models, which jointly optimise all synthesis components, two-stage pipelines separate the acoustic model and vocoder, making it possible to analyze how each stage contributes to synthesis variations.

% two-stage pipelines allow for better isolation of synthesis artefacts, aiding source attribution.

To model real-world synthetic speech generation while maintaining a systematic variation, we use pretrained models from the VCTK dataset~\cite{vctk}. Rather than retraining all models under identical conditions, leveraging existing models ensures representative variability of real-world synthesisers while maintaining strict control over evaluation protocols. We focus on widely used synthesis pipelines, selecting combinations of acoustic and vocoder models that are commonly available online. This approach reflects possible real-world attackers, ensuring that evaluations remain practically relevant.

% To balance control and computational efficiency, we use pretrained models trained on the VCTK dataset~\cite{vctk}. While training all models under identical conditions would be ideal, using existing weights allows for a diverse, realistic comparison. 

We generate speech using 13 tools, each representing a distinct attack—a unique combination or parametrization\footnote{Different parameters (e.g., number of attention heads) or weights.} of acoustic and vocoder models. The dataset follows the VCTK corpus~\cite{vctk} and adopts the ASVspoof2019 Logical Access (LA) data generation protocol~\cite{WANG2020101114}. STOPA 
%Our dataset 
is attack-disjoint from ASVspoof2019 LA, ensuring no direct overlap in synthetic speech generation. While speaker identities overlap between ASVspoof2019’s \textit{dev} set and our trace embedding extraction set (Section~\ref{sec:structure}), this should not introduce data leakage, as speaker identity is not a discriminative feature in source tracing. Consequently, source tracing models pretrained on ASVspoof2019 %pretrained models 
remain compatible with our dataset, provided that attribution focuses on synthesis methods rather than speaker-dependent characteristics.

We collected pretrained model weights from publicly available implementations on GitHub, using only those trained on the VCTK dataset. We then followed the ASVspoof2019 data generation protocols to create seen (\textit{trn}, \textit{dev}) and unseen (\textit{eval}) subsets. %During this process, 
Utterances not included in the pretrained weights or missing from the VCTK dataset were excluded, as detailed in the dataset README. For synthesisers requiring a target speaker utterance (AA11, AA13), we selected the longest available bonafide utterance for that speaker. Since we rely on pretrained weights, we had no control over the original training speakers, making speaker leakage between the \textit{train}, \textit{dev}, and \textit{eval} subsets likely.

% Next, metadata was extracted for all synthesized utterances, and the audio files, protocol files, and metadata were organized into their respective directories.

To manage dataset size while preserving essential variability, we applied a selection process. As part of this process, we included common utterances, having identical linguistic content across TTS and VC attacks, while selecting the remaining utterances at random. Incorporating common utterances enables content-controlled analysis and studies on the impact of identical linguistic content across different attacks. This approach is useful for examining attack consistency, evaluating content-dependent spoofing traits, and, in our case, investigating whether shared utterances enhance attribution performance. The number of common utterances was determined using the elbow criterion on pairwise overlap values, balancing their presence while avoiding excessive overlap between attacks.

\begin{figure}[t]
    \centering
    \includegraphics[width=0.9\linewidth]{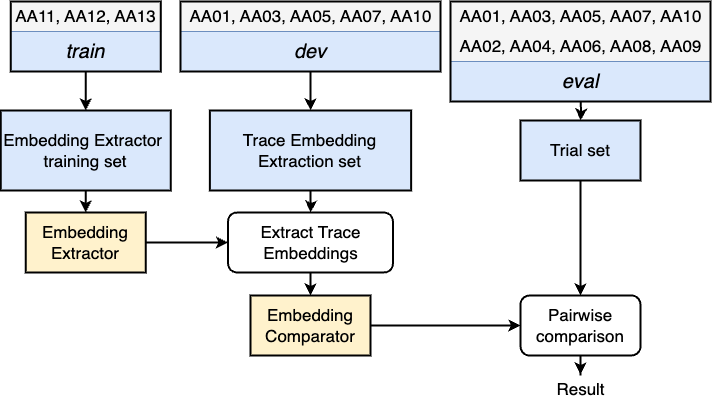}
    \caption{Dataset structure visualisation.}
    \label{fig:stopa-structure}
\end{figure}

\begin{table}[t]
    \centering
    \caption{Count of trials by evaluation conditions. \textit{TE} denotes a trace embedding.}
     \label{tab:trials}
    \resizebox{0.85\columnwidth}{!}{
    \begin{tabular}{@{}ll@{\quad}ll@{}}
    \toprule
    \textbf{Condition} & \textbf{Trials} & \textbf{Condition} & \textbf{Trials} \\ 
    \midrule
    Same attack  & 3,149,000  & Different attack  & 28,341,000 \\
    Same AM      & 6,298,000  & Different AM      & 25,192,000 \\
    Same VM      & 6,298,000  & Different VM      & 25,192,000 \\
    Known TE     & 15,745,000 & Unknown TE        & 15,745,000 \\
    \midrule
    \multicolumn{2}{c}{\textbf{All trials}} & \multicolumn{2}{c}{31,490,000} \\
    \bottomrule
    \end{tabular}
    }
\end{table}

\subsection{Post-processing}

We applied a post-processing pipeline to detect and mitigate shortcut artifacts~\cite{Wang2024}, analyzing the following features: silence durations, peak amplitude, utterance duration, and energy. 
% Silence was identified using voice activity detection (VAD) and root mean square energy (RMSE).

For each waveform $\mathbf{x}$, we removed DC bias and normalized amplitude before detecting non-speech segments. 
Speech onset $s^*$ and offset $e^*$ were determined based on voice activity detection (VAD) activation or root mean square energy (RMSE) exceeding a threshold $\theta_{\text{rmse}}$. 
Leading and trailing silence were identified by counting consecutive non-speech frames, with stable RMSE over $W$ frames signalling the end of the speech. 
Empirical thresholds ($\theta_{\text{rmse}} = 0.05, W = 18$) were optimized to remove non-speech while preserving intelligibility. We evaluated intelligibility using Word Error Rate (WER) with Whisper-large-v3~\cite{whisper} and JiWER~\cite{jiwer}. Samples with WER $>$ 100 (1\%) were removed from \textit{train}/\textit{dev} sets while preserving speaker balance but retained in \textit{eval}, as attribution models should also be able to handle degraded inputs.

\subsection{Detection protocol and evaluation conditions}
\label{sec:structure}

The dataset is organized to support source tracing as a detection task. Its structure follows the original \textit{train}, \textit{dev}, and \textit{eval} subsets, as shown in Figure~\ref{fig:stopa-structure}. The evaluation protocol defines conditions under which comparisons are made, specifying training, reference, evaluation partitions, and the creation of trial pairs.

The training set consists of three attacks and 20 speakers not present in any further sets. The reference set, derived from the original \textit{dev} set, includes varying conditions to evaluate different embedding extraction methods. To systematically assess these methods, trace embedding models vary across two dimensions: (1) \emph{the amount of data used}, ranging from a single utterance per attack to all available utterances, and (2) \emph{the utterance selection strategy}, which either uses common utterances across all attacks or a more realistic scenario where utterances are randomly sampled for each attack.

% Within this set, trace embedding models define variations across two dimensions, ensuring a structured and reproducible evaluation setup:
% \begin{enumerate}
%     \item \textit{Amount of data used} - from single utterance per attack to all available utterances
%     \item \textit{Utterance selection strategy} - either common utterances across all attacks or a more realistic scenario where utterances are randomly sampled for each attack
% \end{enumerate}

Finally, the trial set, derived from the original \textit{eval} subset, defines pairwise comparison for assessing source tracing performance. It includes 10 attack types: five known (present in the reference set) and five unknown. The protocol assigns multiple labels (same attack, same AM, or same VM), enabling analysis at different levels of granularity. The protocol defines 31,490,000 trials, with detailed statistics provided in Table~\ref{tab:trials}.

\subsection{Dataset compatibility with pretrained SSL models}

Self-supervised learning (SSL) models are widely used in deepfake detection, but training data overlap with STOPA must be avoided to prevent evaluation bias. %Most 
Several SSLs, including XLSR-Wav2Vec2~\cite{xlsr-wav2vec}, WavLM~\cite{wavLM}, and Whisper~\cite{whisper}, have no exposure to VCTK, ensuring broad compatibility. 

\begin{table}[t]
    \centering
    \caption{Pooled EER (\%) for used systems across all conditions. AASIST CM denotes the ASVspoof2019-trained countermeasure system.}
    \label{tab:results}
    \resizebox{0.9\columnwidth}{!}{
    \begin{tabular}{l l ccc}
        \toprule
        & \textbf{System} & \textbf{ATK} & \textbf{AM} & \textbf{VM} \\
        \midrule
        \multirow{3}{*}{\shortstack{\textit{Known} \\ \textit{Attacks}}} 
        & ResNet-34   & 47.61 & 47.61 & 47.81 \\
        & AASIST CM & 39.15 & 39.15 & 39.68 \\
        & AASIST STOPA  & 47.05 & 47.05 & 45.25 \\
        \midrule
        \multirow{3}{*}{\shortstack{\textit{Unknown} \\ \textit{Attacks}}} 
        & ResNet-34   & 49.55 & 50.32 & 49.14 \\
        & AASIST CM & 35.34 & 38.69 & 37.31 \\
        & AASIST STOPA  & 47.75 & 49.37 & 48.68 \\
        \bottomrule
    \end{tabular}
    }
\end{table}

\section{Demonstration of STOPA Use}

% To demonstrate the usage of STOPA, we conduct experiments to classify the attack, acoustic model, and vocoder model. We use three embedding extractors: ResNet-34~\cite{dao24_asvspoof}, AASIST~\cite{aasist} (pretrained on ASVspoof2019 LA), and a fine-tuned AASIST trained on STOPA. ResNet-34 is trained, and AASIST fine-tuned for 10 epochs.

While our core contribution is the design and construction of STOPA itself, it is necessary to demonstrate its use in the open-world evaluation setting. Our purpose is \emph{not} to optimize any systems for best performance but to provide pilot results and highlight novel challenges brought in by STOPA. 

We address attack type, acoustic model, and vocoder model classification using three 'trace embedding' extractors: AASIST CM~\cite{aasist} countermeasure (trained on ASVspoof2019 LA for binary spoof detection), and ResNet-34~\cite{dao24_asvspoof} and AASIST STOPA, both trained on the STOPA training partition for attribution. ResNet-34 and AASIST STOPA are trained for 100 epochs. 
% \textbf{We remind the reader that none of these models is trained to classify the attacks, vocoder, or models that we actually use for source tracing; they are trained on completely different sets of classes.} 
We emphasize that none of these models have been exposed to the attacks used in the source tracing reference and trial sets, as the STOPA dataset is explicitly designed to be attack-disjoint. The `zero-shot' mindset (the ability to add new attacks without needing to retrain the embedding extractor) is a key design factor of STOPA protocols.

% Trace embeddings for each attack are extracted based on the predefined models: let the trace embedding for an utterance $x$ be denoted as $t$ and obtained as $t = e(x)$, where $e$ represents the embedding extractor. The trace embedding for each synthesis condition is then computed as a mean embedding: $t_{cond} = \frac{1}{N}\sum^{N}_{i=1}e(x_i)$, where N is the number of utterances given by the source embedding model condition. In evaluation, a test utterance $x_t$ is scored based on its cosine similarity to the known trace embedding $t_{cond}$, computed as: $s = 1 - \frac{t_{cond} \cdot t_{test}}{|t_{cond}| \cdot |t_{test}|}$.

The actual trace embeddings for each attack are extracted using the predefined models. Following the simplest known classifier from speaker detection literature, we form a model for each attack by averaging the training embeddings. %embeddings across all included utterances. 
Test utterances are then compared to each reference using cosine similarity. %Note that neither computing the average embedding nor cosine scoring uses the other attacks in any way.
A \emph{positive} trial is where the ground-truth label (attack, vocoder, or acoustic model ID) in the model and the test match; otherwise, it is a \emph{negative} trial.
%Scores are positive if labels match and negative otherwise. 
The negative set is further split into two subsets: (1) a different known attack and (2) an unknown attack, allowing for EER computation in both scenarios. The latter represents 'the unknown unknown'.

Table~\ref{tab:results} shows error rates for detecting attack type (ATK), acoustic model (AM), and vocoder model (VM). The very high EERs %(chance level is 50\%) 
reflects the difficulty of our open-world task setting. The best performance was obtained with the ASVspoof2019-trained AASIST CM. Both AASIST STOPA and the ResNet-34 models solely trained on STOPA produce, essentially, a chance rate. Since ASVspoof 2019 contains a larger number of training attacks, this strongly suggests that the more limited set of 3 STOPA attacks is not sufficient for capturing 'general variability' of attack-specific traits. Furthermore, using common vs. non-common trace embedding utterances has no impact on final performance.

\begin{figure}[t]
    \centering
    \includegraphics[width=0.9\linewidth]{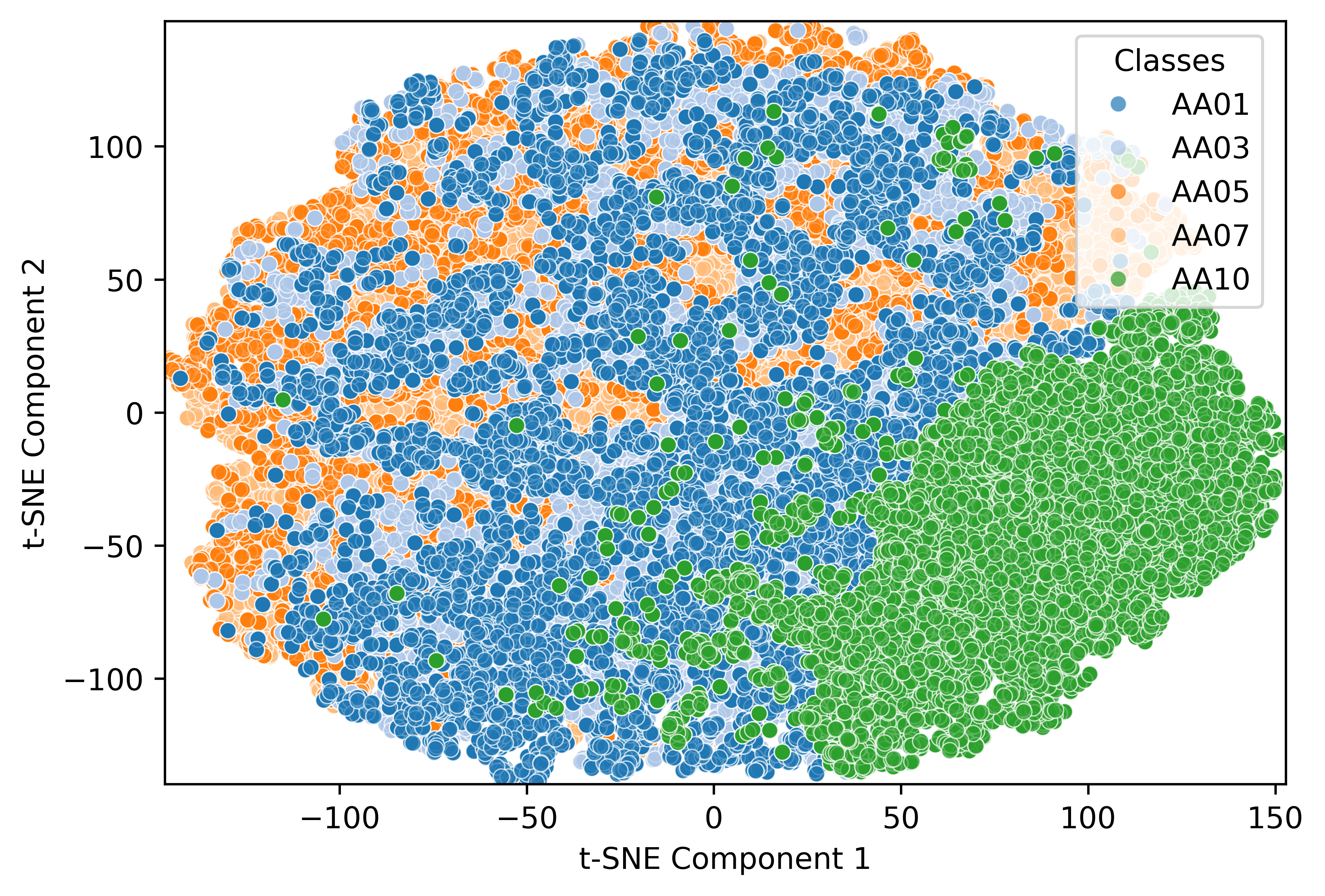}
    \caption{t-SNE trace embedding space visualisation for the ASVspoof2019-pretrained AASIST model.}
    \label{fig:t-sne}
\end{figure}

%High error rates drove 
A further t-SNE analysis (Figure~\ref{fig:t-sne}) reveals a complex embedding space for ASVspoof2019-trained AASIST CM, where only AA10 forms a clear cluster. Training on STOPA did not form any clusters, leading to increased EER. Thus, the mean embedding approach was ineffective due to this extensive cluster overlap.

It might be useful to reflect these results on those typically obtained from speaker detection studies. Modern speaker embedding extractors, such as ECAPA-TDNN, are typically trained using \emph{thousands} of speakers (classes). This leads to a discriminative 'speaker trait' space that transfers easily to classify previously unseen speakers. In contrast, we used only a handful of attacks (3 in STOPA) in training. 

\section{Conclusions}

We introduced STOPA, a novel dataset for synthetic speech source tracing, featuring 13 distinct attacks with systematic variations in acoustic and vocoder models. Unlike existing datasets, STOPA is designed to address 'open-world' source tracing. Inspired by NIST speaker detection benchmarks, our evaluation protocols unify source tracing and unknown attack detection under a scalable framework. By prohibiting the use of other attack data for training or testing, we encourage the development of dynamic 'attack signature databases' that evolve over time, similar to antivirus systems.

Our evaluation shows that established models like AASIST and ResNet fail to structure the signature space, resulting in attack detection EERs exceeding 30\%. This demonstrates the fundamental challenge of zero-shot source tracing and highlights the need for improved embedding extraction and comparison strategies. STOPA thus provides a critical benchmark for advancing scalable and adaptive source tracing systems.

% STOPA sets a strong foundation for further research, offering challenging benchmarks and a new perspective on open-set source tracing.

% This paper presents a novel dataset of synthetic speech for source tracing with systematic variation in acoustic and vocoder models, resulting in 13 attacks with detailed metadata. It addresses critical gaps in source tracing as it provides a baseline for further development in source tracing domain. 

% We outperform the related work because: we provide systematic variation in spoofing samples, we provide detailed metadata for samples and models that created them, we provide specific source tracing protocols for pairwise comparison

% Using this database, we show how to approach source tracing as a detection problem and provide baseline results for further research

% Our findings emphasize that source attribution gets more difficult when multiple attacks share the same architecture and we want to predict the specific attack, opening possibilities for new directions (solutions?) in source tracing

% STOPA helps further research and development in source tracing by providing reliable baseline and different approach to the source tracing task.

% In future verify with tools that we trained to eliminate the speaker leaks.

\section{Acknowledgements}

This work was partially supported by the Brno University of Technology (internal project FIT-S-23-8151) and the Academy of Finland (DecisionNo. 349605, project ”SPEECHFAKES”). The authors wish to acknowledge CSC – IT Center for Science, Finland, for computational resources.

\bibliographystyle{IEEEtran}
\bibliography{mybib}

\end{document}